\documentstyle[aps,prl,psfig,multicol]{revtex}
\begin{document}
\draft

\title{Nonlinear voltage dependence
of  the shot noise in mesoscopic degenerate conductors with strong 
electron-electron
scattering}
\author{E. G. Mishchenko}

\address{Instituut-Lorentz, Universiteit Leiden,
P.O. Box 9506, 2300 RA Leiden, The Netherlands 
and
\\
L.D. Landau Institute for Theoretical Physics,
Russian Academy of Sciences, Kosygin 2, Moscow 117334, Russia}

\maketitle
\begin{abstract}
It is shown that measurements of 
zero-frequency shot noise can 
provide information on electron-electron
interaction, because the strong  interaction
results in the nonlinear voltage dependence of the shot noise
in metallic wires. This 
is due to the fact that
the Wiedemann-Franz law is no longer valid in
the case of considerable electron-electron interaction. 
The deviations from this law increase
the noise power and make it dependent strongly
on the ratio of electron-electron and electron-impurity
scattering rates. 
\end{abstract}
\pacs{PACS numbers: 73.23.-b, 72.70.+m, 72.10-d, 71.10.Ay}
\begin{multicols}{2}

Current  fluctuations 
in nonequilibrium mesoscopic diffusive conductors
proportional to the average current 
$\bar{I}$ manifest the discreteness of charge
carriers and are usually addressed as the shot noise
(for a recent review of the subject see \cite{BB}).
The effect depends on the
 length of the conductor $L$, the electron-impurity  $l_{\rm ei}$ and 
the electron-electron $l_{\rm ee}$ scattering mean free lengths.
In order to observe shot noise, the
electron-phonon mean free length $l_{\rm ep}$ has 
to be the largest scale of
problem. This means that inelastic effects 
of electron-lattice thermalization can be
disregarded. Otherwise, at $l_{\rm ep} < L$,
the noise power 
$S=2\int dt~\overline{\delta I(t) \delta I(0)}$
just approaches the
Johnson-Nyquist equilibrium 
value $4GT$, where $ G$ is the conductance and $T$ is the 
lattice temperature.
\par  Different regimes of shot noise exist. 
\par (i) Diffusive regime, $l_{\rm ei} \ll L \ll \rm l_{ee}$. In this case the effects of 
electron-electron (e-e)
scattering are negligible. 
The energy of each electron  is conserved
during its diffusive motion  
through a 
conductor as soon as the
electron-impurity scattering is elastic. 
The electronic distribution 
function satisfies the diffusion 
equation and has a two-step shape 
$f(\epsilon,x)=(1/2-x/L) f_0(\epsilon-eV/2)
+(1/2+x/L) f_0(\epsilon+eV/2)$, here $f_0$ is the Fermi-Dirac 
function, and $\pm eV/2$ are the shifts of chemical
potentials of the left and right reservoirs, under bias voltage $V$.
The noise power in this regime is known
to be $1/3$ of the Poissonian value $2eI$ \cite{BeB,Nag92}.
\par  (ii) Hot-electron regime, $l_{\rm ei} \ll l_{\rm ee} \ll L$.  
The e-e scattering 
is still small in a sense that all transport processes are governed by 
the impurity scattering. However, e-e 
scattering is already efficient enough to smear  the
two-step partition function and thermalize electrons to the
local
Fermi-Dirac distribution with
some effective temperature profile $T_{\rm e} (x) \sim 0.3 ~ eV \gg T$, 
that is to be found from the equation for energy
transfer. The noise power, now simply representing
the average Johnson-Nyquist noise in  the conductor with 
inhomogeneous temperature distribution 
 $T_{\rm e}(x)$, is equal to $\sqrt{3}/4$ instead of $1/3$
value \cite{Nag95,Koz95}.
The crossover from the 
diffusive to the hot-electron regime has been experimentally
confirmed \cite{SMD,HOS}.
\par
These values of shot noise are {\it universal} \cite{Naz,SL},
i.e.\ they do not depend on 
the applied voltage,
shape of the conductor, anisotropy, distribution and concentration
of impurities, etc. It has been noted  \cite{SL} that 
the universality of the hot-electron shot noise has its origin in
the Wiedemann-Franz law for impurity scattering
(proportionality of the thermal conductivity to
the current conductivity, see Ref.\ \cite{LP}).
\par (iii)  $l_{\rm ee} < l_{\rm ei} \ll 
L$. This regime has not been studied yet. It was pointed out
 that  the electron 
distribution function becomes Fermian
 at relatively weak e-e scattering and does not 
change with its further increase \cite{Nag98}. 
Therefore, in order to obtain information
about e-e scattering 
the finite  frequency noise power has been
studied, which is
sensitive to the details of Coulomb screening and
e-e interaction
\cite{Nag98,NAL}.
\par  However, 
 the zero-frequency noise power
becomes dependent on  the
details of e-e 
scattering if the applied voltage is  high enough. 
Shot noise power in mesoscopic wires
is a fruit  of both the energy transport and the charge transport.
As soon as even 
normal processes of e-e 
scattering affect considerably the thermal conductivity 
(violating the Wiedemann-Franz law) they must 
result in a change of the shot noise power.
 Here we show that this is really the case. 
Despite the fact that 
electron distribution function is of the Fermi-Dirac form 
regardless of the ratio between $l_{\rm ee}$ and $l_{\rm ei}$, the 
temperature profile $T_{\rm e}(x)$ is very sensitive to this ratio.
The shot noise power then acquires nonlinear dependence on the
applied voltage and might even exceed the Poissonian value.
It becomes a possible probe of e-e
interaction, as it happens, e.g., in the case of 
 non-degenerate
mesoscopic conductors (although because of
completely different reason)  \cite{Nag99,SMB,GMP}. 
\par
Super-Poissonian noise has been reported in a number 
of mesoscopic systems. In Refs. \cite{B,ILM} the enhancement
of shot noise was experimentally found in resonant-tunneling
structures biased in the negative differential resistance
regions of their {\it I-V} characteristics. Magnetic field 
has been
used to pronounce this enhancement \cite{KMB}. Coulomb
interaction effects on the shot noise in resonant quantum wells 
near an instability threshold were studied in Ref.\ \cite{BB99}.
The effects of nonlinear {\it I-V} characteristic were considered  microscopically with the self-consistent
Coulomb potential taken into account \cite{WWW}.
In these studies disorder does not play a role. 
\par
In the present paper we show how
super-Poissonian noise can occur in 
 a disordered metallic
wire with strong e-e scattering.
Despite the fact that {\it I-V} characteristic of such a wire 
is still linear, the noise power has a nonlinear voltage 
dependence.
To simplify the problem as much as possible we disregard
all Umklapp 
processes.
As usual, the convenient starting point is the stationary
Boltzmann equation,
\begin{equation}
\label{kin}
{\bf v} \cdot \nabla f_{\bf p}({\bf r}) + e{\bf E}\cdot
\frac{\partial f_{\bf p}({\bf r})}{\partial {\bf p}}=I_{\rm ei}[f_{\bf p}]+
I_{\rm ee}[f_{\bf p}]+ {\cal L}_{\bf p},
\end{equation}
for the electron distribution function in
the phase spase $ f_{\bf p}({\bf r})$,
here $I_{\rm ei}[f_{\bf p}]$ and  $I_{\rm ee}[f_{\bf p}]$ are
 the electron-impurity
and e-e collision integrals respectively, 
\begin{eqnarray}
\label{eicol}
&& I_{\rm ei}[f_{\bf p}] =\int\frac{d^3 p'}{(2\pi)^3}
w_{\rm ei} \delta(\epsilon_{ p}-\epsilon_{p'})
 ( f_{\bf p'}-  f_{\bf p} ),\\  
\label{eecol}
&&I_{\rm ee}[f_{\bf p}] =\int\frac{2d^3 p'}{(2\pi)^3}
\frac{2d^3 k}{(2\pi)^3}
w_{\rm ee} \delta (\epsilon_{p} +\epsilon_{k} -
\epsilon_{p'} - \epsilon_{ k'})\nonumber\\ 
&&\mbox{} \times [f_{\bf p'} f_{\bf k'}
(1-f_{\bf p})(1-f_{\bf k})
-f_{\bf p} f_{\bf k} (1-f_{\bf p'})(1-f_{\bf k'})],
\end{eqnarray}
where ${\bf p}+{\bf p'}={\bf k}+{\bf k'}$.
The quasimomentum conservation law
is exact due to the absence of Umklapp processes.
The scattering amplitudes of the electron-impurity $w_{\rm ei}$
and e-e interaction $w_{\rm ee}$ 
are independent of 
the directions of scattering
particles (this restriction is not crucial).
The last term in Eq.\ (\ref{kin}) is 
the extraneous Langevin source,  zero on average $\overline{
{\cal L}_{\bf p}}=0$, with the correlator 
$\overline{{\cal L}_{\bf p}{\cal L}_{\bf k}}$ defined by the very structure of 
the 
collision integrals (\ref{eicol})-(\ref{eecol}) \cite{Kog69}. Both  electron-impurity and 
e-e collisions contribute to this correlator.
Its exact form is quite cumbersome but will not be needed.
By multiplying the kinetic equation (\ref{kin}) by
the kinetic energy $\epsilon_p-\mu$, and integrating
over the momentum space one gets the
energy balance equation,
\begin{equation}
\label{bal}
\nabla \cdot
 \int \frac{2 d^3 p}{(2\pi)^3} {\bf v}  (\epsilon_{ p} - \mu) f_{\bf p}
({\bf r})
=e {\bf E}  \cdot
 \int \frac{2 d^3 p}{(2\pi)^3} {\bf v}  f_{\bf p}({\bf r}),
\end{equation}
which simply means that the dissipative energy flow is equal to the 
Joule heat.
\par To solve the equation (\ref{kin}) we make the substitution \cite{AKh},
\begin{equation}
\label{sol}
f_{\bf p}({\bf r})=f_0(\xi)+  {\bf v}
\cdot {\bf q} (\xi)\frac{\partial f_0}{\partial \xi}T_{\rm e}^{-1},
\end{equation}
here $f_0(\xi)$ is the Fermi-Dirac function of
the energy
variable $\xi=(\epsilon_p-\mu)/T_{\rm e}$, with
the effective temperature $T_{\rm e}({\bf r})$.
[Throughout the paper we use the units such that 
$\hbar=k_B=1$.]
The (yet unknown) functions $T_e({\bf r})$
and ${\bf q}(\xi)$ determine the nonequilibrium
distribution of electrons near the Fermi surface.
It is convenient to write the function 
${\bf q}(\xi)={\bf q}_{\rm s}(\xi) +{\bf q}_{\rm a}
(\xi)$ as the sum of even
${\bf q}_{\rm s}(-\xi)={\bf q}_{\rm s}(\xi)$ and 
odd function ${\bf q}_{\rm a}(-\xi)=-{\bf q}_{\rm a}(\xi)$ respectively.
In the leading order in $T_{\rm e}/\mu$, 
the symmetric function ${\bf q}_{s}(\xi)$ determines the electric current
while the antisymmetric one ${\bf q}_{a}(\xi)$ gives the dissipative heat flow.
Now the energy balance equation (\ref{bal}) takes 
the form,
\begin{equation}
\label{bal1}
\nabla \cdot
T_{\rm e} \int\limits_{-\infty}^{\infty} d\xi~ \frac{\partial f_0}
{\partial \xi} \xi {\bf q}_{\rm a} (\xi)=e {\bf E} \cdot
 \int\limits_{-\infty}^{\infty} d\xi ~ \frac{\partial f_0}
{\partial \xi} {\bf q}_{\rm s}(\xi),
\end{equation}
as soon as we neglect terms of order  $T_{\rm e}/\mu$.
The integrals  in Eq.\ (\ref{bal1}) are expanded 
over the infinite energy axis 
due to the fast exponential
decay of the factor $\partial f_0/\partial \xi$.
The kinetic equation (\ref{kin}) splits
upon the substitution (\ref{sol}) into two independent
integral 
equations for the even and odd function respectively,
\begin{eqnarray}
\label{even}
 -e{\bf E}&=&\tau_{\rm ei} ^{-1}{\bf q}_{\rm s}(\xi) +
 \int\limits_{-\infty}^
{\infty} d\eta~ K(\xi,\eta)
[{\bf q}_{\rm s}(\xi)
-{\bf q}_{\rm s}(\eta) ],\\
\label{odd}
\xi\nabla T_{\rm e}&=& \tau ^{-1}_{\rm ei} {\bf q}_{\rm a}(\xi)
+ \int\limits_{-\infty}^
{\infty} d\eta~ K(\xi,\eta)
[{\bf q}_{\rm a}(\xi)
-\case{1}{3}{\bf q}_{\rm a}(\eta)].
\end{eqnarray}
here $\tau^{-1}_{\rm ei}= w_{\rm ei}p_F m/2\pi^2$ 
is the electron-impurity
collision rate, and
the kernel function is given by
$$
K(\xi,\eta)=\frac{m^3 w_{\rm ee} T^2_{\rm e}}{2\pi^4}
\frac{(e^{-\xi}+1)(\eta-\xi)}{(e^{-\eta}+1)
(e^{\eta-\xi}-1)}.
$$
To derive the last terms in Eqs.\ (\ref{even})-(\ref{odd}) one has to
perform integrations  in the collision
integral (\ref{eecol})
with respect to the angle and energy 
variables (see Ref.\ \cite{AKh} for
details).
The solution of Eq.\ (\ref{even})
is simply given by the constant,
\begin{equation}
\label{sym}
{\bf q}_{\rm s}(\xi)=-e\tau_{\rm ei}{\bf E}.
\end{equation}
The independence of this solution of the details of e-e
interaction reflects the fact that the electric conductivity in
the absence of Umklapp processes is
determined only by the impurity scattering.
\par The exact solution of Eq.\ (\ref{odd}) is much more 
complicated and can be obtained by the method of Ref.\ \cite{BS},
by which  the thermal conductivity of a clean Fermi liquid
was found.
The brief outline of the method is as follows. By making
the Fourier transform of the integral equation (\ref{odd})
one gets an inhomogeneous second-order differential
equation  for the 
function
${\bf g}(k)=\int d\xi ~
e^{-ik\xi} ~ {\bf q}_{\rm a}(\xi) /(\pi \cosh{[\xi/2]})$,
\begin{eqnarray}
\label{shro}
\frac{d^2 {\bf g}}{dk^2}&+&\left(\frac{2}{3\cosh^2{[\pi k]}}-\gamma
\right){\bf g}\nonumber\\
&& \mbox{} = \frac{8i\pi^4 p_{F}^3}{m^3 w_{\rm ee} T_{\rm e}}
\frac{\sinh{[\pi k]}}{\cosh^2{[\pi k]}} \nabla T_{\rm e},
\end{eqnarray}
where the parameter $\gamma$ of the relative strength of 
e-e
interaction is introduced,
$
\gamma=1+4\pi^2/(\tau_{\rm ei} m^3 w_{\rm ee} T_{\rm e}^2).
$
The function ${\bf g}(k)$ is found by using the 
expansion over the eigenfunctions of 
the homogeneous equation (\ref{shro}), namely
over the Jacobi polynomials $g_{n}=\zeta^{\sqrt{\gamma}/2}
\sqrt{1-\zeta} ~ P_{n}^{(\sqrt{\gamma},1/2)}(1-2\zeta)$, with
$\zeta =1/\cosh^2{[\pi k]}$.
For the thermal 
conductivity, defined as usual as
\begin{equation}
\kappa \nabla 
T_{\rm e}=-\frac{p_F^2}{3m\pi^2}\int\limits_{-\infty}^{\infty}d\xi ~
\frac{\partial f_0}{\partial\xi}\xi {\bf q}_{a}(\xi),
\end{equation}
we get the following expression,
\begin{eqnarray}
\label{kappa} \kappa  (T_{\rm e}) =
\frac{\pi^2 p_F^3}{3m^4 w_{\rm ee} T_{\rm e}}
\sum_{n=0}^{\infty} \frac{2\lambda_n+1/2}{\lambda_n (2\lambda_n+1)-1/3}
\nonumber\\
\mbox{} \times \frac{\Gamma^2(\lambda_n)\Gamma(\lambda_n+1+\sqrt{\gamma}/2)
\Gamma(n+3/2)}{\Gamma^2(\lambda_n+3/2) 
\Gamma(\lambda_n+1/2+\sqrt{\gamma}/2) n!},
\end{eqnarray}
where $\lambda_n=n+\sqrt{\gamma}/2+1/2$, and $\Gamma(x)$ is the
Gamma function.
When the concentration of 
impurities decreases, 
$\gamma \rightarrow 1$, the thermal conductivity
(\ref{kappa}) approaches the known 
expression of the clean limit \cite{BS},
while in the case  of weak e-e
scattering, $\gamma \rightarrow \infty$,
the conventional Wiedemann-Franz law \cite{LP}
is recovered,
\begin{equation}
\label{wid}
\displaystyle
\kappa =\left\{ \begin{array}{ll} \displaystyle
\frac{\pi^2 p_F^3}{3m^4 w_{\rm ee} T_{\rm e}}A, ~~ A=0.78, &
\gamma \rightarrow 1,\\
\displaystyle
\frac{p_F^3\tau_{\rm ei} T_{\rm e}}{9m} \equiv \kappa_0, &
\gamma \rightarrow \infty. \end{array} \right.
\end{equation}
The general expression (\ref{kappa}) 
is the complicated function of
temperature. In what follows, we will use 
the simple interpolation formula built
upon the asymptotic  properties (\ref{wid}),
\begin{equation}
\label{int}
\kappa  (T_{\rm e})
 = \frac{\kappa_0(T_{\rm e})}{1+\beta(
T_{\rm e})},
\end{equation}
where the ratio of the e-e and electron-impurity
scattering rates is defined as
$\beta(T_{\rm e}) \equiv \tau_{\rm ei}/\tau_{\rm ee}(T_{\rm e}) 
= 
\tau_{\rm ei} m^3 w_{\rm ee}T_{\rm e}^2 /3A\pi^2$. Fig.\ 1
demonstrates excellent agreement of this interpolation
formula
with  the exact dependence (\ref{kappa}).
\begin{figure}
  \unitlength 1cm
  \begin{center}
 \begin{picture}(8,5.7)
 \put(-6.8,-13.0){\includegraphics{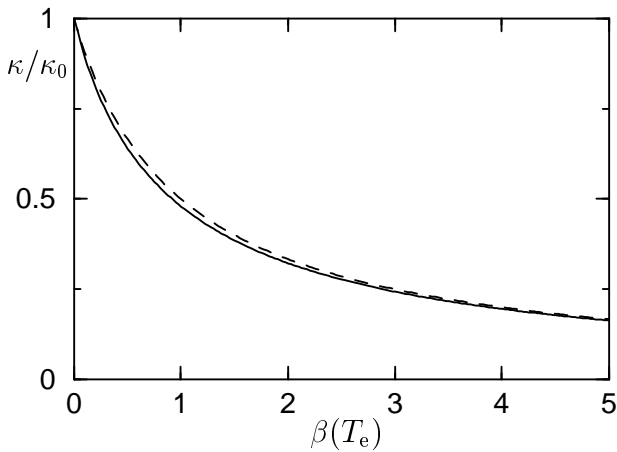}}
  \end{picture}
  \end{center}
\caption{Deviation of the thermal
conductivity (\ref{kappa}) from
the Wiedemann-Franz law plotted versus $\beta(T_{\rm e})=
\tau_{\rm ei}/\tau_{\rm ee}
(T_{\rm e})$. The dashed  line shows the
approximation by the interpolation formula (\ref{int}).}
\end{figure}
\par Substituting Eqs.\ (\ref{sym}) and (\ref{int})
into the energy balance equation (\ref{bal1}), we
obtain the equation for the electron temperature profile,
\begin{equation}
\label{prof}
\frac{d}{dx}\left(\frac{T_{\rm e}}{1+\beta (T_{\rm e})}
\frac{d T_{\rm e}}{dx}   \right)=-\frac{3}{\pi^2}(eE)^2.
\end{equation}
We assume that a conductor of length $L$ is in a contact
with two reservoirs at equilibrium 
with zero temperature $T_{\rm e}(\pm L/2)=0$.
The solution of Eq.\ (\ref{prof}) gives
 the effective temperature profile inside the conductor,
\begin{equation}
\label{profile}
T_{\rm e}(x) = \frac{eV}{\sqrt{\beta(eV)}}
\sqrt{\exp{\bigg[\frac{3\beta(
eV)}{4\pi^2} \left(1-\frac{4x^2}{L^2}\right)\bigg]}-1}.
\end{equation}
To get the expression for the shot noise power it is sufficient
to note that locally (at any given point $x$) the principal
term in the electronic  distribution (\ref{sol})
 is equilibrium-like
and, therefore, the noise is of the Johnson-Nyquist thermal 
type \cite{Nag95,Koz95}. 
The noise power is then given by the averaging of the
thermal noise over the length of the conductor
\cite{com},
\begin{equation}
\label{noise}
S=\frac{4G}{L}\int\limits_{-L/2}^{L/2} dx ~T_{\rm e} (x).
\end{equation}
Substituting Eq.\ (\ref{profile}) into the expression 
(\ref{noise})
we get the shot noise power,
\begin{equation}
\label{p}
{S}={2e\bar{I}} F(V), 
\end{equation}
where the so-called Fano factor (noise suppression factor)
is voltage-dependent and given by the integral
\begin{equation}
\label{fano}
F(V)=\frac{2}{\sqrt{\beta(eV)}}
\int\limits_0^1 dz \sqrt{\exp{\big[\frac{3\beta(eV)}
{4\pi^2} (1-z^2)\big]}-1}.
\end{equation}
Fig.\ 2 shows the Fano factor as a function of 
the e-e collision rate.
\par The shot noise (\ref{p})-(\ref{fano}) is not universal
as it depends nonlinearly on the applied voltage. It also
becomes sensitive to the geometry of the conductor.
The most striking feature of Fig.\ 2 is the monotonous
increase of the shot noise power up to and  above its Poissonian
value $F=1$.
This is different from the prediction, $F \le 1$, of the
quantum linear statistic theory of the shot noise \cite{Les,But}.
This is due to the fact that  the
linear
statistic theory does not include effects of inelastic
scattering.
It is understood however, that in order to observe  
super-Poissonian values,
very strict condition should be satisfied 
$(eV/\mu)^2p_Fl_{\rm ei} \sim 10^2$, making it 
difficult to achieve such a regime.
From the experimental point of view, lower values of $\beta (eV) \le 10$
are of more interest, at which case e-e interaction contributes
a correction to the $\sqrt{3}/4$ noise power,
$F=\sqrt{3}/4(1+9\beta(eV)/64 \pi^2)$.
The main reason why the shot noise power becomes nonlinear under strong e-e 
scattering while the conductance stays Ohmic is due to the
fact that e-e interaction conserves the net momentum of
particles and therefore does not affect the average current 
but 
{\it does} affect the dissipative energy flow that turns
out to be important for current fluctuations.
\begin{figure}
  \unitlength 1cm
  \begin{center}
 \begin{picture}(8,5.7)
 \put(-6.8,-13.0){\includegraphics{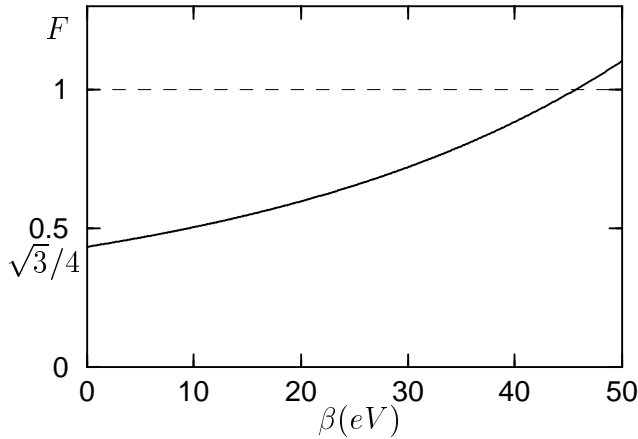}}
  \end{picture}
  \end{center}
\caption{The Fano factor dependence on the effective ratio
of the e-e and electron-impurity scattering 
$\beta (eV)$ at voltage $V$.
The Fano factor starts with $\sqrt{3}/4$ value for weak
e-e scattering and increases monotonously with
its strength.}
\end{figure}
\par The calculation presented here essentially  assumes
that the interference between e-e and impurity scattering
\cite{AA}
is negligible. This is true for sufficiently high
effective electron temperatures (i.e. for high voltages),
$T_{\rm e} \gg \tau^{-1}_{\rm ei}$ \cite{RS}. In order to observe
effects discussed in this paper the mesoscopic conductor has to be 
prepared sufficiently clean to make the ratio $\beta (e V)$
as large as possible. 
\par To summarise, we presented here a situation
where the universality of shot-noise is removed by
a sufficiently strong e-e interaction
and finite voltages. 
The consideration based on the Boltzmann equation and restricted
to the three-dimensional case is given. However, a fully microscopic
theory
for the shot noise in a strongly interacting system theory is 
needed, especially for low dimensional systems.
As to do this is usually not an easy task, the Monte Carlo
simulations could provide helpful insights into the problem.
\par
I gratefully appreciate a number of useful
discussions with C.\ W.\ J. Beenakker and M.\ Titov. This work
was supported by the Dutch Science Foundation NWO/FOM.

\end{multicols}

\begin{references}
\bibitem{BB} Ya. M. Blanter and M. B\"uttiker, cond-mat/9910158,
Phys. Rep. (to be published).
\bibitem{BeB} C. W. J. Beenakker and M. B\"uttiker,
Phys. Rev. B {\bf 46}, 1889 (1992).
\bibitem{Nag92} K. E. Nagaev, Phys.\ Lett.\ A {\bf 169}, 103 (1992).
\bibitem{Nag95} K. E. Nagaev, Phys. Rev. B {\bf 52}, 4740 (1995).
\bibitem{Koz95} V. I. Kozub and A. M. Rudin, Phys. Rev. B {\bf 52},
7853 (1995).
\bibitem{SMD} A. H, Steinbach, J. M. Martinis, and M. H. Devoret,
Phys. Rev. Lett. {\bf 76}, 3806 (1996).
\bibitem{HOS} M. Henny, S. Ober\-hol\-zer, C. Strunk and C. 
Sch\"o\-nen\-ber\-ger,
Phys. Rev. B {59},2871 (1999).
\bibitem{Naz} Yu. V. Nazarov, Phys. Rev. Lett. {\bf 73}, 134 (1994).
\bibitem{SL} E. V. Sukhorukov and D. Loss, Phys. Rev. B {\bf 59},
13054 (1999); Phys. Rev. Lett. {\bf 80}, 4959 (1998).
\bibitem{LP} E. M. Lifshitz and L. P. Pitaevskii, {\it Physical
Kinetics} (Pergamon, Oxford, 1981).
\bibitem{Nag98} K. E. Nagaev, Phys. Rev. B {\bf 58},
R7512 (1998).
\bibitem{NAL} Y. Naveh, D. V. Averin, and K.K. Likharev,
Phys. Rev. B {\bf 59}, 2848 (1999);
Phys. Rev. Lett. {\bf 79}, 3482 (1997).
\bibitem{Nag99} K. E. Nagaev, Phys. Rev. Lett. {\bf 83}, 1267 (1999).
\bibitem{SMB} H. Schomerus, E. G. Mishchenko, and C. W. J. 
Beenakker, Phys. Rev. B {\bf 60}, 5839 (1999);
cond-mat/9907027.
\bibitem{GMP} T. Gonzalez, J. Mateos, D. Pardo, L. Reggiani, and 
O. M. Bulashenko, Physica B
{\bf 272}, 282 (1999).
\bibitem{B} E. R. Brown, IEE Trans. Electron. Devices {\bf 39},
2686 (1992).
\bibitem{ILM} G. Iannacone, G. Lombardi, M. Macucci,
and B. Pellegrini, Phys. Rev. Lett. {\bf 80}, 1054 (1998).
\bibitem{KMB} V. V. Kuznetsov, E. E. Mendez, J. D. Bruno,
and J. T. Pham, Phys. Rev. B {\bf 58}, R10159 (1998);
E. E. Mendez, V. V. Kuznetsov, D. Chokin, and J. D. Bruno,
Physica E {\bf 6}, 335 (2000).
\bibitem{BB99} Ya. M. Blanter and M. B\"uttiker,
Phys. Rev. B {\bf 59}, 10217 (1999).
\bibitem{WWW} Y. Wei, B. Wang, J. Wang, and H. Guo,
Phys. Rev. B {\bf 60}, 16900, (1999).
\bibitem{Kog69} Sh. M. Kogan and A. Ya. Shulman, Zh. Eksp. Teor. Fiz.
{\bf 56}, 862 (1969) [Sov. Phys. JETP {\bf 29}, 467 (1969)].
\bibitem{AKh} A. A. Abrikosov and I. M. Khalatnikov,
Rept. Progr. Phys. {\bf 22}, 329 (1959).
\bibitem{BS} G. A. Brooker and J. Sykes, Phys. Rev. Lett.
{\bf 21}, 279 (1968). 
\bibitem{com} The formula (\ref{noise}) can also be derived
directly from the Boltzmann-Langevin equation (\ref{kin})
with  sources ${\cal L}_{\bf p}$. 
The  normal e-e collisions
do not change the expression (\ref{noise}) for the 
current fluctuations in the same manner
as they do not affect the average conductance.
\bibitem{Les} G. B. Lesovik, JETP Lett. {\bf 49}, 592 (1989).
\bibitem{But} M. B\"uttiker, Phys. Rev. Lett. {\bf 65}, 2901 (1990).
\bibitem{AA} B. L. Altshuler and A. G. Aronov, in {\it Electron-Electron
Interactions in Disordered Systems}, edited by A. L. Efros and M. Pollak
(Elsevier Science Publishers B. V., City, 1985).
\bibitem{RS} J. Rammer and H. Smith,  Rev. Mod. Phys. {\bf 58},
323 (1985).
\end{references}
\end{document}